\documentclass[12pt]{iopart}
\usepackage{amssymb}
\usepackage[dvips]{epsfig}
\begin{document}
\jl{2} \letter{Enhanced dielectronic recombination of lithium-like
Ti$^{19+}$ ions in external $\bi{E}\times\bi{B}$ fields}

\author{T Bartsch\dag , S Schippers\dag\footnote[4]{email: Stefan.E.Schippers@strz.uni-giessen.de},
M Beutelspacher\ddag, S B\"{o}hm\dag,\\ M Grieser\ddag, G
Gwinner\ddag, A A Saghiri\ddag, G Saathoff\ddag,\\ R Schuch\S, D
Schwalm\ddag, A Wolf\ddag~and A M\"uller\dag}

\address{\dag Institut f\"ur Kernphysik, Universit\"at Gie{\ss}en, 35392 Gie{\ss}en, Germany\\
         \ddag Max-Planck-Institut f\"ur Kernphysik and  Physikalisches Institut der Universit\"at Heidelberg, 69117
         Heidelberg, Germany\\
         \S Department of Physics, Stockholm University, 10405 Stockholm, Sweden}

\begin{abstract}

Dielectronic recombination(DR) of lithium-like Ti$^{19+}$($1s^2 2s$) ions via $2s\to 2p$
core excitations has been measured at the Heidelberg heavy ion storage ring TSR. We find
that not only external electric fields ($0 \leq E_y \leq 280$~V/cm) but also crossed
magnetic fields (30~mT$ \leq B_z \leq$ 80~mT) influence the DR via high-$n$
$2p_jn\ell$-Rydberg resonances. This result confirms our previous finding for isoelectronic
Cl$^{14+}$ ions [Bartsch T {\it et al}, \PRL {\bf 82}, 3779 (1999)] that experimentally
established the sensitivity of DR to $\bi{E}\times\bi{B}$ fields. In the present
investigation the larger $2p_{1/2}-2p_{3/2}$ fine structure splitting of Ti$^{19+}$ allowed
us to study separately the influence of external fields via the two series of Rydberg DR
resonances attached to the $2s\to 2p_{1/2}$ and $2s\to 2p_{3/2}$ excitations of the Li-like
core, extracting initial slopes and saturation fields of the enhancement. We find that for
$E_y \gtrsim 80$~V/cm the field induced enhancement is about 1.8 times stronger
for the
$2p_{3/2}$ series than for the $2p_{1/2}$ series.
\end{abstract}


Dielectronic recombination (DR) is a fundamental electron-ion collision process well known
to be important in astrophysical and fusion plasmas (Dubau and Volont\'e 1980). It proceeds
in two steps
\begin{equation}
  e^- + A^{q+} \to [A^{(q-1)+}]^{**} \to A^{(q-1)+}+ h\nu
\label{eq:DR}
\end{equation}
where in the first step the initially free electron is captured into a bound state $n \ell$
of the ion with simultaneous excitation of a core electron. This dielectronic capture (DC,
time inverse of autoionization) can only occur if the energy $E$ of the incident free
electron in the electron-ion center-of-mass (c.m.) frame matches the resonance condition $E
= E_{\rm res} = E_d-E_i$ where $E_i$ and $E_d$ are the total energies of all bound electrons
in the initial and in the doubly excited state, respectively. Employing the principle of
detailed balance the DC cross section can be calculated from the autoionization rate $A_{\rm
a}(d\to i)$ for a transition from the doubly excited state $d$ to the initial state $i$:
\begin{equation}
  \sigma^{\rm (DC)}(E) = S_0 \frac{g_d}{2g_i}\frac{1}{E}
  \frac{A_{\rm a}(d\to i)\Gamma_d}{(E-E_{\rm res})^2+\Gamma_d^2/4}
                   \label{eq:sigmaDC}
\end{equation}
with $S_0=7.88\times10^{-31}$cm$^2$eV$^2$s, statistical weights
$g_d$ and $g_i$ and $\Gamma_d = \hbar\,[\sum_k A_{\rm a}(d\to k) +
\sum_{f'}A_{\rm r}(d\to f')]$ denoting the total width of the
doubly excited state $d$. The summation indices $k$ and $f'$ run
over all states which from $d$ can be either reached by
autoionization or by radiative transitions with rates $A_{\rm
a}(d\to k)$ and $A_{\rm r}(d\to f')$, respectively.

In the second step of reaction \eref{eq:DR} the new charge state
is stabilized by photon emission from the intermediate doubly
excited state, thereby transferring the ion into a final state $f$
below the first ionization limit. This radiative stabilization
competes with autoionization which would transfer the ion back
into its initial charge state with the net effect being resonant
electron scattering. Accordingly, in order to obtain the cross
section for DR one has to multiply the DC cross section from
equation \eref{eq:sigmaDC} by the branching ratio $[\sum_f
A_r(d\to f)]/\Gamma_d$ for radiative stabilization. Integrating
the resulting expression over the c.~m.\ energy and assuming
$\Gamma_d \ll E_{\rm res}$ yields the DR resonance strength due to
the intermediate state $d$ in the isolated resonance approximation
(Shore 1969), i.e.\
\begin{equation}
  \bar{\sigma}_d = S_0 \frac{g_d}{2g_i}\frac{2\pi}{E_{\rm res}}
  \frac{A_{\rm a}(d\to i)\sum_f A_{\rm r}(d\to f)}{\sum_k A_{\rm a}(d\to k) +
  \sum_{f'}A_{\rm r}(d\to f')}.
                   \label{eq:strength}
\end{equation}

In case of $\Delta n=0$ DR of Li-like ions, i.e.\ for $1s^22s\to 1s^22p$ core excitations,
the dominant decay channels of the doubly excited intermediate state (for very high $n$) can
be identified as $2p_jn\ell\to 2sn\ell$ radiative and $2p_jn\ell\to 2sE_{\rm res}\ell'$
autoionizing transitions with rates denoted as $A_{\rm r}$ and $A_{n\ell}$, respectively.
Since the radiative transition only involves the excited core electron, its rate $A_{\rm r}$
to a good approximation is independent of the quantum numbers $n$ and $\ell$ of the excited
Rydberg electron. Neglecting all other transitions such as $2p_{3/2}n\ell \to
2p_{1/2}E'\ell'$ or $2p_jn\ell\to 2p_jn'\ell'$, equation \eref{eq:strength} simplifies to
\begin{equation}
  \bar{\sigma}_{n\ell} = (2j+1)(2\ell+1)S_0 \frac{\pi }{E_n}
  \frac{A_{n\ell}\, A_{\rm r}}{A_{n\ell} + A_{\rm r}}\approx
  (2j+1)(2\ell+1) S_0 \frac{\pi}{E_n} A_<
                   \label{eq:sigmanl}
\end{equation}
using the Rydberg resonance energy $E_{\rm res}= E_n$, $g_d = 2(2j+1)(2\ell+1)$ and $g_i=2$;
in the r.h.s\ approximation $A_<$ denotes the lesser of $A_{\rm r}$ and $A_{n\ell}$.
Autoionizing rates decrease as $\propto n^{-3}$ and even more rapidly with $\ell$, such that
at a given $n$ the relation $A_{n\ell}>A_{\rm r}$ holds only for $n\ell$-Rydberg states with
angular momentum $\ell$ below a limit $\ell_{\rm c}$. Thus $(\ell_{\rm c}+1)^2$ sublevels
dominantly contribute to DR for a given core state $j$ and consequently for the $n$-manifold
$2p_jn\ell$ of doubly excited states the resonance strength is given as
\begin{equation}
  \bar{\sigma}_n \approx (2j+1)[\ell_{\rm c}(n)+1]^2 S_0 \frac{\pi }{E_n} A_{\rm r}.\label{eq:sigman}
\end{equation}
Within this `counting of states' picture the effect of external
electric fields on DR is readily explained. In external electric
fields Stark mixing of high-$\ell$ with low-$\ell$ levels occurs.
This yields autoionization rates which are lower for low-$\ell$
and higher for high-$\ell$ states as compared to the field free
situation. The net effect is an increase of $\ell_{\rm c}$, i.e.\
an increase of the number of states participating in DR. Since
high-$n$ Rydberg states are more easily perturbed by external
electric fields than low-$n$ states the electric field induced
enhancement of DR is stronger for higher-$n$ $2p_jn\ell$ DR
resonances.

This effect of external {\it electric} fields on DR was recognized early by Burgess and
Summers (1969) and Jacobs \etal (1976). Electric field enhancement of DR was subsequently
found in numerous theoretical calculations (Hahn 1997). The first clear experimental
verification of this effect has been given by M\"{u}ller \etal (1986) who investigated DR in the
presence of external fields (DRF) of singly charged Mg$^+$ ions under controlled conditions.
Further DRF experiments with multiply charged C$^{3+}$ ions (Young \etal 1994, Savin \etal
1996) and Si$^{11+}$ ions (Bartsch \etal 1997) also revealed drastic DR rate enhancements by
electric fields. Especially the Si$^{11+}$ experiment which employed merged electron and ion
beams at a heavy ion storage ring equipped with an electron cooler produced results with
unprecedented accuracy, enabling a detailed comparison with theory. Whereas the overall
agreement between experiment and theory as for the magnitude of the effect was fair,
discrepancies remained in the functional dependence of the rate enhancement on the electric
field strength (Bartsch \etal 1997). This finding stimulated theoretical investigations of
the role of the additional magnetic field which in storage ring DR experiments is always
present, since it guides and confines the electron beam within the electron cooler. In a
model calculation Robicheaux and Pindzola (1997) found that in a configuration of crossed
$\bi{E}$ and $\bi{B}$ fields indeed the magnetic field through the mixing of $m$ levels
influences the rate enhancement generated by the electric field through the mixing of $\ell$
levels. More detailed calculations (Griffin \etal 1998a, Robicheaux \etal 1998) confirmed
these results. It should be noted that in theoretical calculations by Huber and Bottcher
(1980) no influence of a pure magnetic field ($\bi{E}=0$) on DR was found up to at least
$\bi{B}=5$~T.

Inspired by these predictions we previously performed storage ring DRF experiments using
Li-like Cl$^{14+}$ ions and crossed $\bi{E}$ and $\bi{B}$ fields (Bartsch \etal 1999) where
we clearly discovered a distinct effect of the {\it magnetic} field strength on the
magnitude of the $\bi{E}$-field enhanced DR rate. Shortly after that Klimenko and coworkers
(1999) experimentally verified that for the $m$-mixing to occur the crossed $\bi{E}$ and
$\bi{B}$ arrangement is essential. For the case of parallel $\bi{B}$ and $\bi{E}$ fields,
where $m$ remains a good quantum number, they did not observe any influence of the magnetic
field on the measured recombination signal.

The aim of the present investigation with Li-like Ti$^{19+}$ is to confirm the novel
$\bi{E}\times\bi{B}$ field effect on DR for a heavier Li-like ion. Because of the strong
scaling of the fine-structure splitting with the nuclear charge the
Ti$^{18+}$($1s^22p_{1/2}n\ell$) and Ti$^{18+}$($1s^22p_{3/2}n\ell$) Rydberg series of DR
resonances are well separated in energy. The corresponding series limits occur at 40.12~eV
and 47.81~eV, respectively (Hinnov \etal 1989). This energy difference is large enough that,
in contrast to the Cl$^{14+}$ experiment, here our experimental resolution permits to study
the effect of external fields on both Ti$^{18+}$($1s^22p_jn\ell$) Rydberg series
individually.

The experiments were carried out at the heavy ion storage ring TSR
of the Max-Planck-Institut f\"{u}r Kernphysik in Heidelberg. Here we
only give a brief account of the experimental procedure for DRF
measurements. Details will be given in a forthcoming publication
by Schippers \etal (2000, and references therein) on DRF
measurements with lithium-like Ni$^{25+}$ ions.

Beams of $^{48}$Ti$^{19+}$ ions with intensities up to almost 80~$\mu$A were stored in the
ring at energies of 4.6~MeV/u. The ion beams were cooled by interaction with a
velocity-matched cold beam of electrons which was confined by a magnetic field $\bi{B}$; the
direction of $\bi{B}$ defines that of the electron beam. The electron beam diameter was
30~mm, while that of the cooled ion beam was of the order of 2~mm. First, as in the standard
tuning procedure of the electron cooler, the electron beam was steered so that, along the
straight interaction region of 1.5~m length, the ion beam travelled on the electron beam
center line and the guiding field $\bi{B}$ pointed exactly along the ion beam; this
minimized the electric field in the frame of the ions originating from space charge and
motional ($\bi{v} \times \bi{B}$) fields. A reasonably `electric-field free' measurement of
the DR rate coefficient (with an estimated residual field of at most $\pm 10$\,V/cm) could
then be obtained at high energy resolution by switching the energy of the electrons in the
cooler to different values. The energy range thus covered in the center-of-mass frame
includes all Ti$^{19+}$($1s^2 2p_{j}n\ell$) $\Delta n=0$ DR resonances due to the $2s\to
2p_j$ core excitations. Recombined Ti$^{18+}$ ions were magnetically separated from the
parent Ti$^{19+}$ beam and detected with an efficiency $\geq 95$\% downbeam from the cooler
behind the first bending magnet.

Controlled motional electric fields in the frame of the ions were then applied by
superimposing in the interaction region a defined transverse (horizontal) magnetic field
$B_x \ll B_z$ in addition to the unchanged longitudinal field $B_z$ along the ion beam
direction ($z$). This field was generated by the electron-beam steering coils along the
complete straight section of the electron cooler and created a motional electric field
$E_y=vB_x$ in the frame of the ions at a beam velocity $v$; the total magnetic field
strength $(B_x^2+B_z^2)^{1/2}$, however, remained almost unchanged. Progressively different
electric fields were produced by varying the transverse magnetic field strength $B_x$. At a
given transverse field $B_x$ the electron beam (following the magnetic field lines) and the
ion beam are misaligned by the small angle $B_x/B_z$ so that the distance of the ion beam
from the center of the electron beam varies along the interaction region. This leads to
unwanted effects due to the electron space charge: (I) a variation of the (temperature
average) relative velocity between electrons and ions along the interaction path, resulting
in a degraded energy resolution; (II) creation of an additional electric field $E_x$ which,
in contrast to the imposed field $E_y$, varies along the interaction region. Low electron
densities were chosen in order to keep these effects small. With electron currents of 20~mA,
measurements were performed at an electron density of $6.2\times10^6$\,cm$^{-3}$. The cooler
was operated at longitudinal field strengths $B_z$= 30, 41.8, 60.0 and 80.1~mT. Transverse
fields of $|B_x| \le 0.7$\,mT (measured with an uncertainty of $\pm3$\%) were applied,
corresponding to controlled motional electric fields $|E_y|$ up to 280~V/cm; in all
measurements the misalignment angle was kept below $|B_x/B_z|\lesssim0.02$. The ratio
$E_x/E_y$ of the unwanted electric space charge field and the applied motional field is
expected to vary linearly along the interaction region with $|E_x/E_y|$ remaining always
below 0.07 for all measurements with different experimental parameters.

Before each energy scan with an imposed electric field $E_y$, ions were injected into the
ring, accumulated and then cooled for 1\,s. After that, the cathode potential of the cooler
was offset from cooling by about 1~kV (corresponding to 55~eV in the center-of-mass frame)
and then, the steering coils were set to produce a defined transverse magnetic field $B_x$.
Next, the center-of-mass energy was ramped down from about 55~eV to 1~eV within 4\,s thus
completing a first mini-cycle. After new ion injection and cooling (at $B_x=0$), the next
magnetic steering field $B_x$ (i.e., next $E_y$) was automatically set and a new energy scan
started. The mini-cycles, covering one complete energy scan each, were repeated for a set of
pre-chosen magnetic steering fields. A grand cycle through typically 11 values of $E_y=vB_x$
thus took about 2 minutes and such cycles were repeated until a satisfying level of
statistical uncertainty (below 3\% per channel) had been reached.

\begin{figure}
\begin{center}
 \epsfxsize=12cm \epsfbox{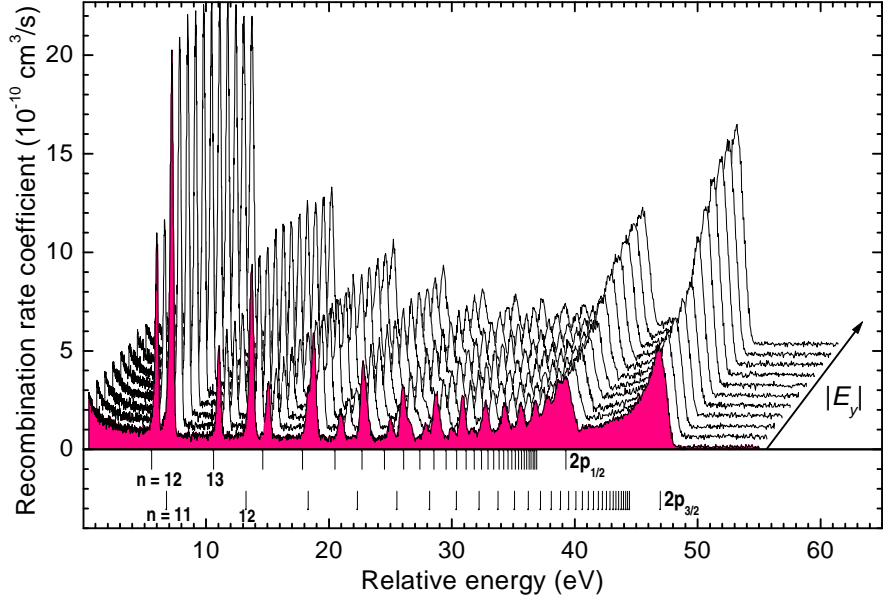}
\end{center}
\caption[]{Absolute recombination rate coefficients measured for
  4.6~Mev/u Ti$^{19+}$ ions at applied motional electric fields
  $|E_y|$ increasing nearly linearly from 0 to 265~V/cm;
  longitudinal magnetic field $B_z$=69~mT, electron density $6
  \times 10^6$~cm$^{-3}$. Energetic positions of the $2p_{1/2}\,n\ell$
  and $2p_{3/2}\,n\ell$ resonances according to the Rydberg formula are
  indicated.
 \label{Ti19spectra} }
\end{figure}

Sets of recombination rate measurements were made for different
longitudinal fields $B_z$.  Using measured beam currents the
spectra were calibrated, reaching an uncertainty of $\pm20\%$ for
absolute and $\pm5\%$ for relative rate coefficients. The
center-of-mass energies were determined (with uncertainties below
$\pm1$\%) from the average relative velocities of electrons and
ions, accounting for the angle between the electron and the ion
beam due to the applied transverse field $B_x$.

A typical set of measurements is presented in
figure~\ref{Ti19spectra} and shows the two series of Ryd\-berg
resonances converging to the $2p_{1/2}$ and $2p_{3/2}$ core
excitation limits.  A significant enhancement of the rate
coefficient with increasing electric field $E_y$ is observed for
high Rydberg states $n \gtrsim 27$, while for the lower-lying
resonances the rate coefficient remains constant.

\begin{figure}
\begin{center}
\epsfxsize=13cm \epsfbox{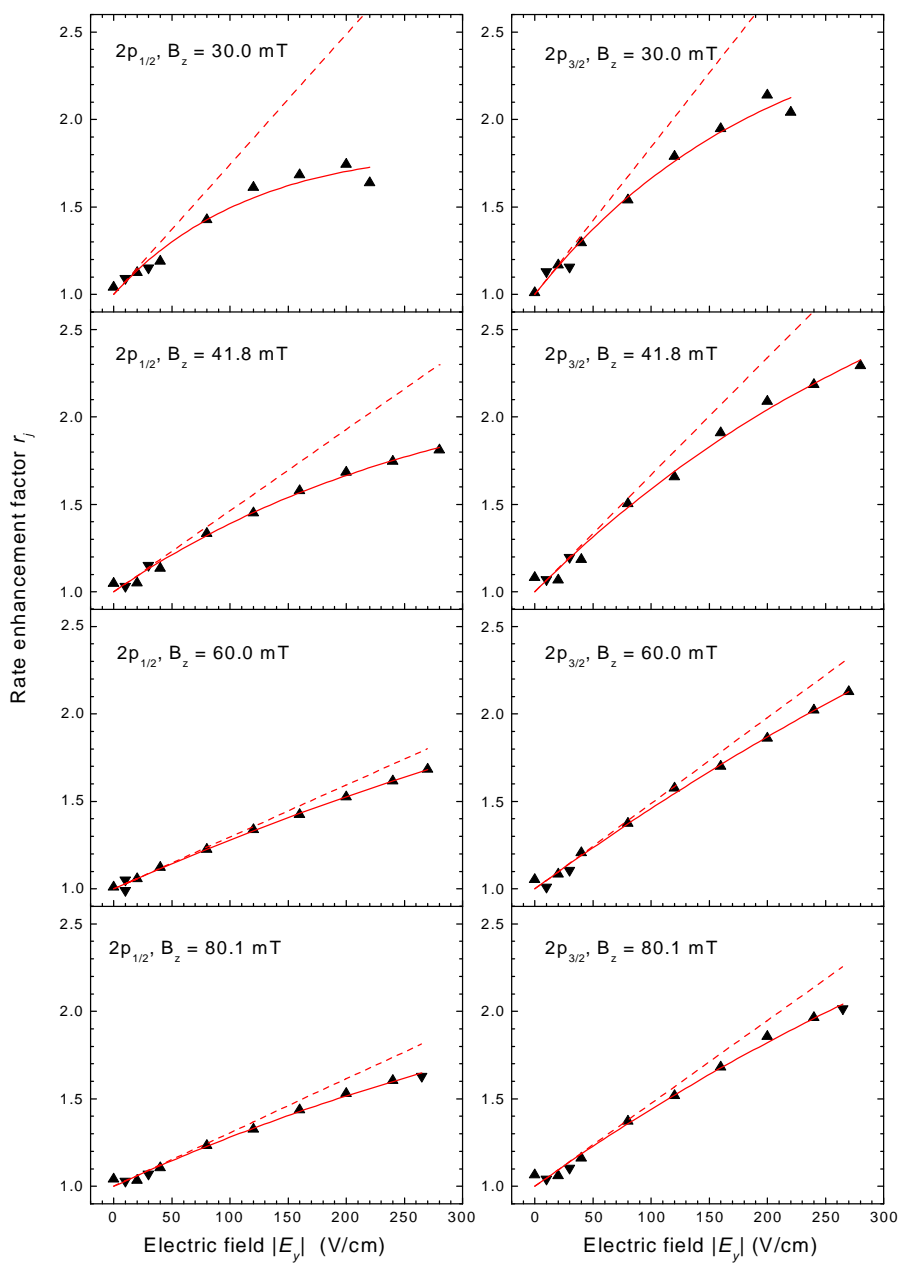}
\end{center}
\caption[]{Measured field enhancement factors (cf.\ equation \protect\eref{area}) $r_{1/2}$
(left) and $r_{3/2}$ (right) as a function of the applied motional electric field strength
$|E_y|$ for different longitudinal magnetic field strengths $B_z = 30.0, 41.8, 60.0$ and
$80.1$~mT (from top to bottom). Triangles pointing up (down) mark data points which have
been measured with positive (negative) $E_y$. The full lines have been fitted to the data
points (cf.\ equation \protect\eref{eq:fit}). The dashed straight lines are tangents to the
fit at $E_y=0$.
 \label{Ti19rj} }
\end{figure}

The enhancement of the DR via high Rydberg states is quantified by extracting rate
coefficients integrated over different energy regions of the measured spectra. The integrals
$I_{1/2}(E_y,B_z)$ and $I_{3/2}(E_y,B_z)$ extend over the energy ranges 33.4--40.57~eV and
40.57--50~eV, respectively, and represent the high-Rydberg contributions of the
$2p_{1/2}n\ell$ and $2p_{3/2}n\ell$ series of Ryd\-berg resonances with $n \geq 27$. For
normalization purposes we also monitor the integral $I_0$ (integration range 4--24~eV)
comprising DR contributions from lower $n$. It should be noted that the maximum quantum
number of Rydberg resonances contributing to the measured recombination rate is limited by
field ionization in the charge analyzing dipole magnet. Taking into account also radiative
decay of high Rydberg states on the way from the cooler to the dipole magnet, we estimate
the maximum quantum number to be $n_c = 115$.

The high-Rydberg contributions $I_{j}(E_y,B_z)$ ($j=1/2,3/2$)
monotonically increase with $|E_y|$, while the lower-$n$
contribution $I_{0}(E_y,B_z)$ remains constant.  In order to
provide a quantity for the following discussion that is
independent of the normalization of the individual spectra, we
consider ratios $I_{j}/I_{0}$ of the high-$n$ to the low-$n$
contribution in a single DR spectrum.  The electric-field
enhancement factor
\begin{equation}
r_j(E_y,B_z) = C_j(B_z)\frac{I_{j}(E_y,B_z)}{I_{0}(E_y,B_z)}
\label{area}
\end{equation}
then directly measures the influence of the external electric field on the DR rates via high
Rydberg states. The constants $C_j(B_z)$ have been chosen such that fits to the data points
(see below) yield $r^{\rm (fit)}_j(0,B_z)=1.0$.

The field enhancement factors $r_{1/2}(E_y,B_z)$ and $r_{3/2}(E_y,B_z)$ found for different
$B_z$ are shown in figure~\ref{Ti19rj} as a function of $|E_y|$. The enhancement factors
turn out to be independent of the sign of $E_y$, as expected. The formula
\begin{equation}
r^{\rm (fit)}_j(E_y,B_z) = 1+s_j(B_z)E_j(B_z)\left\{1-\exp[-E_y/E_j(B_z)]\right\}
\label{eq:fit}
\end{equation}
which we have fitted to the measured enhancement factors, provides a useful parameterization
of our data. The parameters which have been varied during the fits (at fixed values of
$B_z$) are the saturation field $E_j(B_z)$ and the initial slope $s_j(B_z)$; tangents to
$r^{\rm (fit)}$ at $E_y=0$, representing the initial slopes are also displayed in figure
\ref{Ti19rj}. We note that the $E_y = 0$ data points are slightly above the fitted lines.
This is due to the fact that zero applied field $E_y=0$ still implies a residual electric
field $\lesssim10$\,V/cm so that the measured dependence of $r_j(E_y,B_z)$ near $E_y=0$ is
washed out to some extent. At higher electrical field strengths the measured data points
drop below the (dashed) straight lines (cf. figure \ref{Ti19rj}). This is an indication that
the electric field effect is subject to saturation which occurs at higher electric field
strength where the mixing of $\ell$-levels is complete. Presently, higher electric fields
are not accessible in our DR experiment. The fit parameter $E_j(B_z)$ indicates how fast the
saturation regime will be reached.

\begin{figure}
\begin{center}
 \epsfxsize=12cm \epsfbox{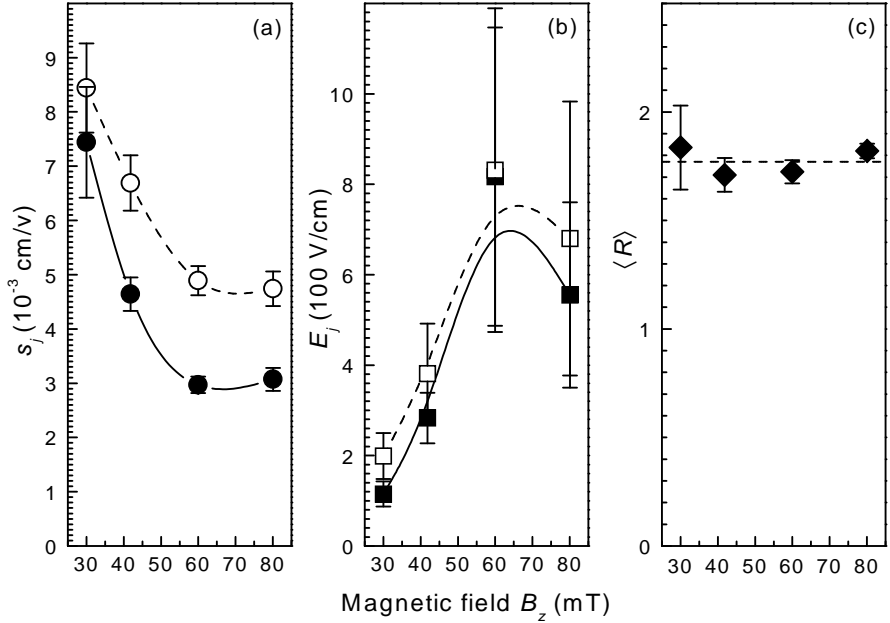}
\end{center}
\caption[]{Dependence of the fit parameters (cf.\ equation
\protect\eref{eq:fit}) $s_{1/2}$ (closed circles), $s_{3/2}$ (open
circles), $E_{1/2}$ (closed squares) and $E_{3/2}$ (open squares)
on the longitudinal magnetic field strength $B_z$. The error bars
were obtained from the fits. The lines are drawn to guide the eye.
The diamonds in panel (c) represent ratios $R$ (cf.\ equation
\protect\eref{eq:R}) averaged over the electric field interval
80~V/cm$\leq \vert E_y \vert \leq$~280~V/cm as a function of
$B_z$. The error bars correspond to one standard deviation. The
dashed straight line represents the mean value $1.77\pm 0.06$.
\label{Ti19sj}}
\end{figure}

The values for the parameters $s_j(B_z)$ and $E_j(B_z)$, which along with their
uncertainties have been obtained from the fits, are displayed in figures \ref{Ti19sj}a and
\ref{Ti19sj}b, respectively, as a function of the magnetic field $B_z$. Both $s_j$ and $E_j$
exhibit a strong dependence on the strength of the magnetic guiding field. The slopes $s_j$
decrease with increasing magnetic field both for the $2p_{1/2}$ and $2p_{3/2}$ series of
Rydberg resonances. This confirms our recent finding for Cl$^{14+}$ ions (Bartsch \etal
1999) where the existence of a sensitivity of DR to external {\it magnetic} fields in an
$\bi{E}\times\bi{B}$ field configuration was experimentally demonstrated for the first time.
The parameters $E_j$ increase with increasing magnetic field strength, i.e.\ at higher $B_z$
the saturation regime will be reached at higher electric fields $\vert E_y\vert$.

While the parameters $E_j$ are not markedly different for the $2p_{1/2}$ and $2p_{3/2}$
series, the slopes $s_{3/2}$ are steeper than the slopes $s_{1/2}$ (open and closed circles
in figure \ref{Ti19sj}a, respectively), i.e.\ the relative increase of the DR line strength
is stronger for the $2p_{3/2}$ series of Rydberg resonances than for the $2p_{1/2}$ series.
For a comparison to recent theoretical predictions (Griffin \etal 1998a, 1998b) we consider
the ratio
\begin{equation}
R(E_y,B_z) = \frac{I_{3/2}(E_y,B_z)-I_{3/2}(0,B_z)}{I_{1/2}(E_y,B_z)-I_{1/2}(0,B_z)}
\label{eq:R}
\end{equation}
of absolute $E_y$ induced DR rate enhancements for the $2p_{3/2}$ and $2p_{1/2}$ series of
Rydberg resonances, which is practically independent of the integration ranges used for the
determination of $I_{1/2}$ and $I_{3/2}$ as long as they cover nearly all DR resonances
affected by the external fields. As a function of $E_y$ the ratio $R$ rises up to $\vert
E_y\vert =80$~V/cm and then essentially stays constant at higher electric fields. Values
$\langle R\rangle$ averaged over the interval 80~V/cm$\leq \vert E_y \vert \leq$~280~V/cm
are plotted in figure \ref{Ti19sj}c which shows that
$\langle R\rangle = 1.77\pm 0.06$
independent of $B_z$.

In view of the fact that for a given $n$ the manifold of $2p_{3/2}n\ell$ resonances contains
twice as many sublevels that can be mixed by external fields as the manifold of
$2p_{1/2}n\ell$ resonances ($8n^2$ vs.\ $4n^2$, cf.\ equation \eref{eq:sigmanl}) one expects
a value of 2 for the ratio $\langle R\rangle$. We here observe a ratio somewhat lower than 2
similar to the value $\langle R\rangle\sim 1.5$ found in our experiments with lithium-like
Ni$^{25+}$ ions (Schippers \etal 2000). In calculations for lithium-like Si$^{11+}$ ions
(Griffin \etal 1998a) and C$^{3+}$ ions (Griffin \etal 1998b) ratios even less than 1 have
been found. This has been attributed to the electrostatic quadrupole-quadrupole interaction
between the $2p$ and the $n\ell$ Rydberg electron in the intermediate doubly excited state,
which more effectively lifts the degeneracy between the $2p_{3/2}n\ell$ than between the
$2p_{1/2}n\ell$ levels. Our experimental results suggest that this effect might be weaker
than theoretically predicted.

Our data emphasize the relevance of the effect of small magnetic fields on DR via high
Rydberg levels in conjunction with the well-known electric-field enhancement. This result
bears important implications upon the charge state balance of ions in astrophysical and
laboratory plasmas where both, electric and magnetic fields are ubiquitous.

We gratefully acknowledge support by BMBF, Bonn, through contracts
No.\ 06 GI 848 and No.\ 06 HD 854 and by the HCM Program of the
European Community.

\section*{References}

\begin{harvard}

\item[] Bartsch T, M\"{u}ller A, Spies W, Linkemann J, Danared H, DeWitt D R, Gao H , Zong W , Schuch R,
Wolf A, Dunn G H, Pindzola M S and Griffin D C 1997 \PRL {\bf 79}
2233

\item[] Bartsch T, Schippers S, M\"{u}ller A, Brandau C, Gwinner G,
Saghiri A A, Beutelspacher M, Grieser M, Schwalm D, Wolf A,
Danared H and Dunn G H 1999 \PRL {\bf 82} 3779


\item[] Burgess A and Summers H P 1969 {\it Astrophys.\ J.} {\bf 157} 1007

\item[] Dubau J and Volont\'e S 1980 \RPP {\bf 43} 199

\item[] Griffin D C, Robicheaux F and Pindzola M S 1998a \PR A {\bf 57} 2798

\item[] Griffin D C, Mitnik D, Pindzola M S and Robicheaux F 1998b \PR A {\bf 58}
4548

\item[] Hahn Y 1997 \RPP{\bf 60} 691

\item Hinnov E and the TFTR operating team, Denne B and the JET operating team 1989 \PR A {\bf 40} 4357

\item[] Huber W A and Bottcher C 1980 \JPB {\bf 13} L399

\item[] Jacobs V L, Davies J and Kepple P C 1976 \PRL {\bf 37} 1390



\item[] V. Klimenko, L. Ko, and T. F. Gallagher 1999 \PRL {\bf 83}
3808

\item[] M\"{u}ller A, Beli\'c D S, DePaola B D, Djuri\'c N, Dunn G H, Mueller D W and Timmer C
1986 \PRL {\bf 56} 127

\item[] Robicheaux F and  Pindzola M S 1997 \PRL {\bf 79} 2237

\item[] Robicheaux F, Pindzola M S, and Griffin D C 1998 \PRL {\bf 80} 1402

\item[] Savin D W, Gardner L D, Reisenfeld D B, Young A R, and Kohl J L 1996 \PR
A {\bf 53} 280

\item Schippers S, Bartsch T, Brandau C, M\"{u}ller A, Gwinner G, Wissler G, Beutelspacher M,
Grieser M, Wolf A and Phaneuf R 2000 to be published 

\item[] Shore B W 1969 {\it Astrophys.\ J.} {\bf 158} 1205

\item[] Young A R, Gardner L D, Savin D W, Lafyatis G P, Chutjian A, Bliman S and Kohl J L 1994 \PR A {\bf 49} 357
\end{harvard}

\end{document}